\documentclass[a4paper]{jpconf}
\usepackage{dcolumn}
\usepackage{bm}
\usepackage[applemac]{inputenc}
\usepackage[spanish,english]{babel}
\usepackage{amsmath,amssymb,amsfonts,latexsym,cancel}
\usepackage{graphicx}
\usepackage{color}
\usepackage{hyperref}

\newcommand*{\Scale}[2][4]{\scalebox{#1}{$#2$}}%
\newcommand{\be}{\begin{equation}}
\newcommand{\ee}{\end{equation}}
\newcommand{\bea}{\begin{eqnarray}}
\newcommand{\eea}{\end{eqnarray}}

\begin{document}
\title{On the renormalization of ultraviolet divergences in the inflationary angular power spectrum}

\author{Adrian del Rio$^1$ and Jose Navarro-Salas$^2$ }

\address{Department of Theoretical Physics, IFIC. A mixed center University of Valencia - CSIC. \\ Faculty of Physics, University of Valencia, Burjassot 46100, Valencia, Spain.}

\ead{$^1$adrian.rio@uv.es, $^2$jnavarro@ific.uv.es}

\begin{abstract}

We revise the role that ultraviolet divergences of quantum fields play in slow-roll inflation, and discuss the  renormalization of cosmological observables from a space-time perspective, namely the angular power spectrum. We first derive an explicit expression for the multipole coefficients $C_{\ell}$ in the Sachs-Wolfe regime in terms of the two-point function of primordial perturbations.   We  then analyze the  ultraviolet behavior, and  point out that the  standard result in the literature is equivalent to a renormalization of $C_{\ell}$ at zero ``adiabatic'' order. We  further argue that  renormalization at second ``adiabatic'' order may be  more appropriate   from the viewpoint of standard quantum field theory. This may change significantly the predictions for $C_{\ell}$, while maintaining scale invariance.

\end{abstract}

\section{Introduction and motivation}

The theory of quantum fields interacting with gravity \cite{parker-toms, birrell-davies},  applied to the very early and rapidly expanding universe, explains well the pattern of temperature anisotropies of the cosmic microwave background (CMB) as well as the large scale structure (LSS) of the universe.  The assumption of a slow-roll inflationary universe \cite{kolb-turner, Weinberg}, in particular, is   useful to account for the huge and detailed cosmological data observed today \cite{planck13}.  In this framework, particle creation \cite{parker66, parker68} is the   fundamental  mechanism driving primordial perturbations that eventually seeded  the tiny fluctuations in the temperature of the CMB. It can also be regarded as the source for the gravitational clumping of matter that gave rise to galaxies and structure formation at late times.

In a curved space-time, several ultraviolet (UV) divergences  arise in the computation of vacuum expectation values, and these infinities can not be removed by  standard methods in Minkowski space-time. Specific methods to define regularization and renormalization in expanding universes have been constructed to account for the new UV divergences sourced by gravitational fields \cite{parker-toms, birrell-davies} (for more recent works, see \cite{landete}).

For definiteness, let $\varphi$ represent a generic free scalar field living in a homogeneous and isotropic {\it Fridmann-Lemaitre-Robertson-Walker} (FLRW) spacetime, with line element  $ds^2=dt^2-a^2(t)d\vec{x}^2$. The field $\varphi$ can be used to describe  scalar (or tensor) perturbations during inflation. In the quantum theory, the free field operator is most generally studied by its expansion in Fourier $k$-modes $\varphi_k(t)$,
 \be
\varphi(t, \vec{x})= \int d^3k \left[A_{\vec{k}}\varphi_k(t)+ A^{\dagger}_{-\vec{k}}\varphi^*_k(t)\right] e^{i\vec{k}\vec{x}} \ ,
\label{eq: field in RW}
\ee
where $A^{\dagger}_{\vec{k}}$ and $A_{\vec{k}}$   are creation and annihilation operators, respectively, which satisfy canonical commutation relations: $[A_{\vec{k}}, A_{\vec{k}\, '}]=0 \ , [A_{\vec{k}}^{\dagger}, A_{\vec{k} \, '}^{\dagger}]=0$ and $[A_{\vec{k}}, A_{\vec{k}\, '}^{\dagger}]= \delta^{3}(\vec k -\vec k\, ')$. As usual, the vacuum state of the quantum field is defined by   $A_{\vec{k}} |0 \rangle= 0$, and excited states in the theory containing quanta are obtained by acting repeatedly with the creation operator. For free fields, the whole theory is entirely determined by specification of  the two-point function:
\bea \label{2pf}\langle  \varphi(t, \vec{x}) \varphi(t, \vec{x}')\rangle =\int d^3k |\varphi_k(t)|^2 e^{i\vec k(\vec x -\vec x')} 
 \ , \eea
 where we have restricted to equal times $t=t'$ for later purposes. 
 We can formally construct local physical observables from the above two-point function.  For instance, the local quantum fluctuations of 
$\varphi$ can be quantified by the mean square fluctuation in the vacuum state
\be  \label{varcoincidence} \langle  \varphi^2(t, \vec{x}) \rangle= \int d^3k |\varphi_k(t)|^2 
 \ . \ee
In either case, it is quite common in cosmology to refer the quantity  $\Delta^2_{\varphi}(k, t)\equiv 4\pi k^3  |\varphi_k(t)|^2$ as the  power spectrum. 

A proper definition of the physical power spectrum in inflationary cosmology is not free of subtleties, as first pointed out in \cite{parker07}, and subsequently studied in  \cite{grg09, glenz-parker, prl09}. In momentum-space, and for a single mode $k$, the power spectrum $\Delta^2_{\varphi}(k, t)$ is well defined.  However, the formal variance  $\langle \varphi^2(\vec{x}, t)  \rangle$, which is an infinite sum of all modes,  diverges in the ultraviolet. There is no doubt that this self-correlator needs renormalization when it is used  to quantify the amplitude of quantum perturbations at a single space-time point, as in (\ref{varcoincidence}). However,  the two-point function (\ref{2pf}) does not need   renormalization when used to quantify physical observables involving correlations. This is because the two-point function has a well-defined definition in the distributional sense, and as such there is no mathematical need for any regularization \cite{Bastero-Gil}, provided we integrate it with test functions of compact support. However, as the spatial points get close together, the two-point function grows without bound, and can produce divergences in physical observables if the functions involved are not of compact support \cite{paper1} (see expression (\ref{cl}) below). 
In \cite{parker07, grg09, glenz-parker, prl09, paper1} it was argued that the physical power spectrum should be defined in terms of renormalized quantities.  The challenging of this proposal for inflationary cosmology and quantum gravity has been recently stressed in  \cite{woodard}. The purpose of this work is to reanalyze these issues, specially from a spacetime viewpoint.

As we have said, (\ref{varcoincidence}) is UV divergent and needs to be renormalized according to  standard rules,
\be \label{varrenormalized} \langle  \varphi^2(t, \vec{x}) \rangle= \lim_{\vec x \to \vec x'}[ \langle  \varphi(t, \vec{x}) \varphi(t, \vec{x}')\rangle - {^{(N)}}G_{Ad} ((t, \vec x), (t, \vec x'))]  \ , \ee
where ${^{(N)}}G_{Ad} ((t, \vec x), (t, \vec x'))$ is the  $N$th order adiabatic subtraction term. The subtraction terms can be specified via the adiabatic regularization method, or, equivalently, using the DeWitt-Schwinger scheme (for more details see \cite{adrian-pepe}.) To properly cancel the UV divergences in our $4$-dimensional spacetime, the second adiabatic order $N=2$ is the right answer for the mean square fluctuation $\label{var} \langle  \varphi^2(t, \vec{x}) \rangle$. However,  the proper adiabatic order of the subtraction term depends on the particular  physical quantity  to evaluate. For instance, the computation of the renormalized expectation value of the stress-energy tensor $\langle T_{\mu\nu}(t, \vec x)\rangle $ needs subtraction up to the fourth adiabatic order, using $[\langle  \varphi(t, \vec{x}) \varphi(t, \vec{x}')\rangle - {^{(4)}}G_{Ad} ((t, \vec x), (t, \vec x'))]$ instead of $\langle  \varphi(t, \vec{x}) \varphi(t, \vec{x}')\rangle$, and taking the coincident limit. 

Apart from these fundamental observables, one can also be interested in integrated quantities that can be obtained from the two-point function. An example is given by
\be \label{psp}\langle \varphi_{\vec p} \varphi_{\vec p'} \rangle  = \int d^3 \vec x d^3 \vec x'  e^{i(\vec p \cdot \vec x +\vec p \, '\cdot \vec x \, ' )} \langle  \varphi(t, \vec{x}) \varphi(t, \vec{x}\, ')\rangle=  
|\varphi_k(t)|^2  \delta^3 (\vec p + \vec p \, ')\ , \ee
Similarly, one can also consider, assuming rotational invariance,  the following observable
\bea 
\label{cl} C_{\ell}^{\varphi\varphi} & = & \frac{1}{4\pi}  \int d^2\hat n  \, d^2\hat n' P_{\ell}(\hat n \cdot \hat n')  \langle  \varphi(t, \vec{x}) \varphi(t, \vec{x}')\rangle \nonumber\\
& = &  2\pi \int_{-1}^1 d\cos \theta P_{\ell}(\cos \theta)  \langle  \varphi(t, \vec{x}) \varphi(t, \vec{x}')\rangle 
   =16 \pi^2 \int_0^{\infty} \frac{dk}{k} |\varphi_k(t)|^2 j_{\ell}^2(k|\vec x|)\ ,    
\eea
 where $P_{\ell}$ are the Legendre Polynomials,  and $\cos \theta = \vec n \cdot \vec n'$ is the angle formed by the two spatial directions $\vec{n}=\vec{x}/|\vec{x}|$ and $\vec{n}'=\vec{x}'/|\vec{x}|$. The expression above is, precisely, the angular power spectrum that is employed in cosmology to study the CMB. For completeness, we have also added the equivalent and more familiar expression in momentum space, which involves the spherical Bessel functions $j_{\ell}$.
  Although the former integral (\ref{psp}) is UV finite and no renormalization in needed, the latter expression actually suffers from an UV divergence. Namely, a detailed inspection reveals that the divergence is of  $N=0$ adiabatic order. This conclusion can be deduced from the short-distance (adiabatic) expansion ($\theta \to 0$) of the two point-function:
\be \langle  \varphi(t, \vec{x}) \varphi(t, \vec{x}')\rangle \sim \frac{1}{1-\cos \theta } - \frac{(\frac{1}{6}-\xi)R}{2}\log (1-\cos \theta) + ... \ee
and taking into account that $P_{\ell}(\cos \theta) \sim 1$. Therefore, while (\ref{psp}) should be kept unaltered, expression (\ref{cl}) must be  modified  according to the standard renormalization prescription in quantum field theory. Despite that $\langle  \varphi(t, \vec{x}) \varphi(t, \vec{x}')\rangle$ is well-defined as a distribution, the observable $C_\ell$ is not, because the Legendre Polynomials do not have compact support (i.e., they are not  test functions).  Similarly to what is usually done for $\langle T_{\mu\nu}\rangle$, to obtain a physically sensible quantity for $C_{\ell}^{\varphi\varphi}$, one should then replace $\langle  \varphi(t, \vec{x}) \varphi(t, \vec{x}')\rangle$ by $[ \langle  \varphi(t, \vec{x}) \varphi(t, \vec{x}')\rangle - {^{(0)}}G_{Ad} ((t, \vec x), (t, \vec x'))]$ in expression (\ref{cl}) to guarantee  the UV finiteness of the integral.  The full calculation is displayed in \cite{paper1}.
 
We recall that the appropriate choice for the  adiabatic subtraction order $n$  depends on the physically relevant object. In  cosmology of the CMB, the direct physical observables are temperature correlations, which are  linked to the space-time two-point function  $\langle  \varphi(t, \vec{x}) \varphi(t, \vec{x}')\rangle$. For instance, in the Sachs-Wolfe regime we have $\langle \Delta T (\vec n)\Delta T(\vec n')\rangle_{SW}= \frac{T_0^2}{25} \langle  \varphi(t, \vec{x}) \varphi(t, \vec{x}')\rangle$, where $\varphi (t, \vec x)\equiv {\cal{R}}(t, \vec x)$ is the comoving curvature perturbation. These temperature correlations are the ones upon which other observables like $C^{TT}_{\ell}$ are constructed
\be \label{CL} C_{\ell}^{TT}= \int_{-1}^1 d\cos \theta P_{\ell}(\cos \theta) \langle \Delta T (\vec n)\Delta T(\vec n')\rangle \ . \ee 
Since physical correlations are direct observables that can be measured in an experiment, it seems natural to demand that they must always be finite, even at coincidence $\vec x=\vec x'$. To achieve that, one should then relate $\langle \Delta T (\vec n)\Delta T(\vec n')\rangle$ with the  quantity  $[\langle  \varphi(t, \vec{x}) \varphi(t, \vec{x}')\rangle - {^{(2)}}G_{Ad} ((t, \vec x), (t, \vec x'))]$ to ensure  UV finiteness at coincidence. Therefore, the actual related multipole coefficients should be constructed with second order adiabatic subtractions
\be \label{clr} C^{N=2}_{\ell} =   2\pi \int_{-1}^1 d\cos \theta P_{\ell}(\cos \theta) [ \langle  \varphi(t, \vec{x}) \varphi(t, \vec{x}')\rangle - {^{(2)}}G_{Ad} ((t, \vec x), (t, \vec x'))] \ .  \ee

 \section{Spacetime correlators in slow-roll inflation}

 Let us now focus on  scalar perturbations during slow-roll inflation. For simplicity we will also consider the Sachs-Wolfe regime, for which the transfer functions are trivial. Scalar fluctuations are described through the comoving curvature perturbation field ${\cal{R}} (t, \vec x)$. For single-field inflation, the modes  ${\cal{R}}_k( t)$   defining the Bunch-Davies type vacuum, are  \cite{Weinberg} 
  \be 
 {\cal{R}}_k( t) = \sqrt{\frac{-\pi \eta}{4(2\pi)^3 z^2}} H^{(1)}_{\nu}(-\eta k) \label{modes}\ , 
   \ee 
where $H\equiv \dot a/a$ is the Hubble rate, $H_{\nu}^{(1)}$ is the Hankel function with $\nu= \frac{3}{2} + \frac{2\epsilon +\delta}{1-\epsilon}$, $\epsilon \equiv -\dot{H}/H^2 \ll 1$, and $\delta \equiv \ddot H/2H\dot H$ is a second slow-roll parameter. In this expression we introduced the so-called proper time $\eta$, defined by $d\eta=\frac{dt}{a(t)}$. Moreover, $z\equiv a\dot{\phi}_0/H$, where $\phi_0(t)$ is the homogeneous part of the inflaton field, responsible for the period of inflation itself. 
 
 Using the modes (\ref{modes}) one can work out analytically  the corresponding  two-point function $\langle {\cal R} (t, \vec {x}), {\cal R} (t, \vec {x}') \rangle$. It  is given by
\bea
 \label{2pfR}\langle {\cal R} (t, \vec {x}) {\cal R} (t, \vec {x}') \rangle  = \frac{1}{16 \pi^2 z^2\eta^2} \Gamma\left(\frac{3}{2}+\nu\right) \Gamma \left(\frac{3}{2}-\nu\right) {_{2}}F_1\left(\frac{3}{2}+\nu, \frac{3}{2}-\nu; 2; 1-\frac{(\Delta x)^2}{4\eta^2}\right)  \ ,  
\eea
where ${_{2}}F_1$ is the hypergeometric function, and $\Gamma$ is the gamma function. The two-point separation is described by $\Delta x \equiv |\Delta \vec x|= 2^{\frac{1}{2}}|\vec{x}|(1-\cos \theta)^{1/2}$.
For very short separations we get
\bea
 \langle {\cal R} (t, \vec {x}) {\cal R} (t, \vec {x}') \rangle =  \frac{G H^2(1-\epsilon)^2}{4\pi \epsilon} \left\{  \frac{4}{\Delta \bar{x}^2} +\left[\frac{1}{4}-\nu^2\right]\ \log \frac{\Delta \bar{x}^2}{4} + O(\Delta \bar{x}^0)\right\}   \label{modesbajox}\, ,
\eea
where we defined $\Delta \bar{x} =\Delta x \, a\, H (1-\epsilon)$. From this expression it is clear  that, as expected, the scalar two-point function diverges when $\theta \to 0$. This produces an ultraviolet divergence in the (unrenormalized) expression for multipole coefficients in the Sachs-Wolfe regime, for which one makes the identification $
 \langle \Delta T (\vec n)\Delta T(\vec n')\rangle_{SW}= \frac{T_0^2}{25} \langle  {\cal R}(t, \vec{x}) {\cal R}(t, \vec{x}')\rangle$, 
\be \label{clSWur} C^{SW}_{\ell} =   \frac{2\pi T_0^2}{25} \int_{-1}^1 d\cos \theta P_{\ell}(\cos \theta) \langle  {\cal R} (t, \vec{x}) {\cal R} (t, \vec{x}')\rangle = \infty \ ,  \ee
where $|\vec x| =|\vec x'| = r_L$ is the comoving radial coordinate of the last scattering surface and $\theta$ is the angle formed by  $\vec x$ and $\vec x'$. One can resolve this divergence with the  replacement $\langle  {\cal R} (t, \vec{x}) {\cal R} (t, \vec{x}')\rangle \to [\langle  {\cal R} (t, \vec{x}) {\cal R} (t, \vec{x}')\rangle- G_{Ad}^{(0)}((t, \vec x), (t, \vec x'))]$.  Then, one gets
 \be \label{clSWr0} C^{SW(n=0)}_{\ell} =   \frac{2\pi T_0^2}{25} \int_{-1}^1 d\cos \theta P_{\ell}(\cos \theta) [\langle  {\cal R} (t, \vec{x}) {\cal R} (t, \vec{x}')\rangle -G_{Ad}^{(0)}((t, \vec x), (t, \vec x'))] < \infty \ .  \ee
Evaluating this well-defined integral at late times yields the following result [we define the new parameter $\bar r_L = r_L  \, a\, H (1-\epsilon)$]

\bea \label{ClSWn0}C^{SW(N=0)}_{\ell} =  \frac{8\pi T_0^2}{25}\frac{4\pi G}{\epsilon} \frac{H^2(1-\epsilon)^2}{16\pi^2}\frac{\Gamma(3-n)\Gamma(\ell+\frac{n-1}{2})}{\Gamma(\ell+2-\frac{n-1}{2})} \bar r_L^{1-n}\ , \nonumber \\ \label{19b}
\eea
 where we have used 
$\nu-\frac{3}{2}= \frac{1-n}{2}$, 
and $n$ represents the scalar index of inflation $n=1-4\epsilon-2\delta+O(\epsilon,\delta)^2$. This expression is in exact agreement with the well-known standard result that is obtained via the  momentum-space power spectrum.
It is worth to remark that the same result can be obtained if one  consider the large-scale behavior [i.e. the late-time one during inflation $a|\vec{x}-\vec{x}'| \gg H^{-1}$] of the two-point function \be
 \label{2pfR2}\langle {\cal R} (t, \vec {x}) {\cal R} (t, \vec {x}') \rangle \sim \frac{\Gamma(\frac{3}{2}-\nu) }{4 \pi^2 z^2\eta^2} \frac{\Gamma (\nu)}{\sqrt{\pi}} \left (\frac{\Delta x}{-\eta}\right )^{2(\nu - 3/2)} \ . 
\ee 
and use this expression in (\ref{clSWur}). This is indeed the standard approach, as reported in many textbooks \cite{Weinberg}. Therefore, it is actually equivalent to the calculation of $C^{SW(N=0)}_{\ell}$ with the renormalized two-point function at zero  adiabatic order.

Even though this procedure provides a finite result, what we have really encountered in (\ref{modesbajox}) is the typical quadratic and logarithmic  short-distance behavior of a quantum field in a curved background. So, as argued before, it seems natural to remove both divergences, not only the leading one.  Accordingly, it is natural to propose the following identification
 \bea
 \langle \Delta T (\vec n)\Delta T(\vec n')\rangle_{SW}= \frac{T_0^2}{25} [\langle {\cal R} (t, \vec {x}) {\cal R} (t, \vec {x}\, ')  \rangle-G_{Ad}^{(2)}((t, \vec x), (t, \vec x'))] \ , 
  \label{44}
 \eea 
instead of   $\langle \Delta T (\vec n)\Delta T(\vec n')\rangle_{SW}= \frac{T_0^2}{25} [\langle {\cal R} (t, \vec {x}) {\cal R} (t, \vec {x}\, ')  \rangle-G_{Ad}^{(0)}((t, \vec x), (t, \vec x'))]$, as used in (\ref{clSWr0}) to obtain (\ref{ClSWn0}). 

The subtraction term at second adiabatic order is found to be 
 \small
 \bea
G_{Ad}^{(2)}((t, \vec x), (t, \vec x')) & = &  \frac{G H^2(1-\epsilon)^2}{4\pi \epsilon}\left\{   \frac{4}{\Delta\bar{x}^2}+\left(\frac{1}{4}-\nu^2\right) \log{\frac{\Delta\bar{x}^2}{4}}  \right. \nonumber \\
 & & \left.+\frac{2-\epsilon}{3(1-\epsilon)^2}+\left(\frac{1}{4}-\nu^2\right) \left[2\gamma+\log{\frac{\mu^2}{H^2(1-\epsilon)^2}}\right] \right\}  \label{59}
\ , \eea 
\normalsize
where $\mu$ is a renormalization scale, and $\gamma$ is the Euler constant.
Now we should obtain a suitable expression for the two-point function. A good way to deal with this is to expand  expression (\ref{2pfR}) as a power series of the ``slow-roll'' parameter $\nu$ around $\nu=3/2$, and stay at  first order (for details see  \cite{paper1}). The result reads:
\bea
\Scale[1]{
 \langle {\cal R} (t, \vec x) {\cal R} (t, \vec x') \rangle \approx \frac{G H^2(1-\epsilon)^2}{4\pi \epsilon}\left\{      \frac{4}{\Delta \bar{x}^2} -1   +  \frac{2}{(3/2 - \nu)} \Big(\frac{\Delta\bar x^2}{4}  \Big)^{\nu-3/2}    \right\}   }  \label{approx}
 \eea 
We can now proceed to do the subtraction. The renormalized two-point function  then reads 
\bea
 \langle {\cal R} (t, \vec x) {\cal R} (t, \vec x') \rangle -  G_{Ad}^{(2)}((t, \vec x), (t, \vec x')) & \approx & \frac{G H^2(1-\epsilon)^2}{4\pi \epsilon} \left\{  \frac{2}{(3/2 - \nu)}\Big(\frac{\Delta\bar x^2}{4}  \Big)^{\nu-3/2} +2\log \Delta \bar x^2  \right. \nonumber\\
&&\left. -\frac{5}{3}+4\gamma+2\log{\frac{\mu^2}{H^2}}  \right\} \ . \label{77}  
\eea

We now proceed to compute  the corresponding angular power spectrum from the renormalized two-point function  and for the Sachs-Wolfe regime
\bea
C_{\ell}^{SW(N=2)}=\frac{2\pi T_0^2}{25} \int_{-1}^{1}d\cos \theta P_{\ell}(\cos \theta) [\langle  {\cal R} (t, \vec{x}) {\cal R} (t, \vec{x}')\rangle -G_{Ad}^{(2)}((t, \vec x), (t, \vec x'))]  \ . \label{61}
\eea
To evaluate the logarithmic contributions of  (\ref{77}) to (\ref{61}) we take into account that $\int_{-1}^{1}dy \log(1-y) P_{\ell}(y)= - 2/\ell(\ell+1), \ell = 1, 2, \dots \ $.
The final result  is very well approximated by the following analytical expression:
\bea
C_{\ell}^{SW(N=2)} \approx \frac{4\pi G}{\epsilon} \frac{8\pi T_0^2}{25} \frac{H^2(1-\epsilon)^2\bar r_L^{1-n}}{16\pi^2}  \left\{   \frac{\Gamma(\ell+\frac{n-1}{2})}{\Gamma(\ell+2-\frac{n-1}{2})}-\frac{\bar r_L^{n-1}}{\ell(\ell+1)}\right\}      \ . 	\label{78}
\eea 
The above expression is valid for $\ell\geq 1$, as for $\ell=0$ there would be present all the constant contributions  from the renormalized two-point function (\ref{77}), including the one depending on the renormalization scale. In fact, the renormalization scale may be fixed by imposing the natural condition $C^{SW(N=2)}_{0}=0$. 

Notice that the first term in (\ref{78}) comes form the renormalization procedure at zero adiabatic order, and  reproduces the standard result (\ref{19b}). The second one comes from the purely second adiabatic order subtraction term, but it shows scale invariance as well. Therefore, Eq. (\ref{78}) is also consistent with observations
\cite{planck13}. 
The first term is time-independent, while the second term depends slightly on time through the quantity ${\bar r_L}^{n-1}\equiv \alpha$. This parameter can be regarded as a phenomenological parameter, varying in the range $1 > \alpha > 0$. In the limiting case $\alpha  \to 0$ ones recovers the standard prediction, and this would correspond to the subtraction terms  evaluated after the end of inflation. If the subtraction terms are evaluated a few $e$-foldings after the horizon exit of the scale $r_L$, the parameter $\alpha$ approaches $1$ and the physical significance of the correction increases.

\section{Conclusions}

We have briefly reported a critical   overview on the renormalization of cosmological observables in slow-roll inflation, focusing in particular on the angular power spectrum of the CMB.  To do so, we have analyzed two-point correlators of quantum fields and self-correlators   of scalar primordial perturbations in  quasi-de Sitter spacetime backgrounds.  We have   reanalyzed the evaluation of the multipole coefficients $C_{\ell}$ in terms of the two-point correlator. We have pointed out that the  standard result in the literature for the angular power spectrum is equivalent to renormalizing  $C_{\ell}$ at zero adiabatic order in quantum field theory. We have argued, though, that  renormalization at second adiabatic order is more suitable  from the viewpoint of quantum field theory. This may change significantly the predictions of inflation, provided  the renormalization subtraction terms are evaluated a few $e$-foldings after the first horizon crossing of the scale $r_L$.

\ack
This work is supported by the Research Project of the Spanish MINECO FIS2011-29813-C02-02, and the Consolider Program No. CPANPHY-1205388. A. del Rio is supported by the Spanish Education Ministry Ph.D. fellowship FPU13/04948.

\end{document}